\documentclass[nologo,11pt,a4paper]{ETHpaper}
\usepackage{graphicx, amsmath, amssymb,color,wasysym}
\usepackage{bm}
\usepackage[square,numbers,sort&compress]{natbib}

\begin{document}

\newcommand{\mean}[1]{\left\langle #1 \right\rangle} 
\newcommand{\abs}[1]{\left| #1 \right|} 
\newcommand{\ul}[1]{\underline{#1}}
\renewcommand{\epsilon}{\varepsilon} 
\newcommand{\eps}{\varepsilon} 
\renewcommand*{\=}{{\kern0.1em=\kern0.1em}}
\renewcommand*{\-}{{\kern0.1em-\kern0.1em}} 
\newcommand*{\+}{{\kern0.1em+\kern0.1em}}

\newcommand{\RA}{\Rightarrow}
\newcommand{\bbox}[1]{\mbox{\boldmath $#1$}}

\title{An agent-based framework of active matter with applications in biological and social systems}

\titlealternative{An agent-based framework of active matter with applications in biological and social systems}

\author{Frank Schweitzer}
\authoralternative{Frank Schweitzer}

\address{Chair of Systems Design, ETH Zurich, Weinbergstrasse 58, 8092 Zurich, Switzerland}

\reference{European Journal of Physics (Revised submission)}

\www{\url{http://www.sg.ethz.ch}}

\makeframing
\maketitle

\begin{abstract}
  Active matter, as other types of self-organizing systems, relies on the take-up of energy that can be used for different activities, such as active motion or structure formation. 
  Here we provide an agent-based framework to model these processes at different levels of organization, physical, biological and social, using the same dynamic approach.
  Driving variables describe the take-up, storage and conversion of energy, whereas driven variables describe the energy consuming activities.
  The stochastic dynamics of  both types of variables follow a modified Langevin equation. Additional non-linear functions allow to encode system-specific hypotheses about the relation between driving and driven variables.
  To demonstrate the applicability of this framework, we recast a number of existing models of Brownian agents and Active Brownian Particles. Specifically, active motion, clustering and self-wiring of networks based on chemotactic interactions, online communication and polarization of opinions based on emotional influence are discussed. 
  The framework allows to obtain critical parameters for active motion and the emergence of collective phenomena.
This highlights the role of energy take-up and dissipation in obtaining different dynamic regimes.

  \end{abstract}
\date{\today}

\section{Introduction}

The term \emph{Active Matter} \citep{marchetti2013hydrodynamics,julicher2018hydrodynamic,ramaswamy2010mechanics} was recently introduced to describe the dynamics of systems with the ability to take up energy from the environment.
This energy can be transformed into active motion of the system elements, which is the most studied case \citep{Romanczuk2012a,Ebeling1999}.
But also complex interactions such as chemotaxis \citep{fs-lsg-94,Ebeling2003,Romanczuk2015}, clustering \citep{zottl2016emergent,Grossmann2015} or chiral pattern formation \citep{furthauer2013active} have been investigated.

It is worth noticing that the term \emph{active matter} is \emph{not} a synonym of \emph{living matter}.
The focus of most publications is indeed on biological applications, 
but also the active (self-propelled) motion of artificial particles (micro swimmers) has been experimentally realized (see Table 1 in \citep{bechinger2016active}).

The literature features two different approaches for modeling active matter, one dealing with  macroscopic equations, the other one with particle-based methods.
This publication  shall mainly contribute to the second strand.
Therefore, it is advisable to contrast these two approaches in the beginning.
We thus start with some general reflections.

\paragraph{Some historical remarks.}

To put the recent research about \emph{active matter} into perspective, it is useful to remind on some relations to existing scientific paradigms.
The \emph{systemic} capabilities of active matter to develop and to maintain coherent structures, or collective states, based on the input and the conversion of energy were previously described using the term \emph{self-organization}.
It was heuristically defined as the ``spontaneous 
formation, evolution and differentiation of complex order structures forming in non-linear dynamic systems by way of feedback mechanisms involving the elements of the systems, when these systems have passed a critical distance from the statical equilibrium as a result of the influx of unspecific energy, matter or information.'' \citep{fs-ed-97}. 
The \emph{physics of self-organization and evolution} \citep{ebeling2011physics} has made fundamental contributions to the scientific understanding of self-organizing systems -- the thermodynamics of irreversible processes, deterministic and stochastic models of nonlinear dynamics, the theory of reaction-diffusion processes, information-theoretical approaches to sequences, to name just a few. 

While \emph{self-organization} could be seen as the leading paradigm of the 1970th and 1980th, it was gradually replaced by \emph{complex systems} as the leading paradigm after the year 1990. 
This development came along with a shifting focus. 
Self-organization and pattern formation could be well described on the macroscopic or \emph{systemic} level, for example by coupled partial differential equations. 
With complex systems, i.e. systems composed of a large number of (strongly) interacting elements, the focus was more on the \emph{emergence} of systemic properties from these interactions. 
Hence, the methodological framework to address this process turned towards \emph{agent-based modeling}, where agents represent the system elements \citep{schweitzer2007brownian}.
They have their own internal degrees of freedom and follow their own dynamics. 

The latest leading paradigm, \emph{complex networks} which became prominent after the year 2000, can be seen as a special case of the complex systems paradigm in that it \emph{decomposes} all interactions between agents into \emph{dyadic interactions}, i.e. interactions between \emph{pairs of agents}. 
The complex network then consists of \emph{nodes} representing agents and \emph{links} representing dyadic interations.
With this, the focus shifted again: away from the \emph{agents} and their internal dynamics (which became dots, now) and towards the \emph{topology} and the dynamics of the \emph{network} composed by the lines between dots. 
This abstraction leaves out a lot of important details about the system elements,
to favor generality and universality of the network.
Hence, active matter is one of those application areas where the complex network approach \emph{cannot} be readily applied. 

We have sketched this development to highlight that the natural sciences, in particular statistical physics, have addressed the \emph{potentiality} of matter to become active in the sense described above for quite a long time. 
So, active matter is in fact a self-organizing system, and the existing theoretical concepts to model such systems can be applied and should be rediscovered, if needed.

\paragraph{Different modeling approaches.}

As the discussion also makes clear, there are different methodological approaches to describe active matter -- the \emph{systemic approach} largely builds on phenomenological macroscopic equations, while the \emph{agent-based approach} tries to understand those macroscopic properties based on interactions on the micro,  or agent, level.  

Both of these approaches have their justification dependent on the application area. 
For many experiments in soft matter physics and biophysics, a \emph{continuum model} is the most appropriate way to explain the system dynamics, e.g. turbulence in bacteria or structure formation in the cytoskeleton \citep{Toner2005,marchetti2013hydrodynamics,juelicher2007active}.
Other experiments, on the other hand, require to take a \emph{particle based} or agent-based approach, i.e. their explanation depends on modeling the details of the particles' \emph{motion} and their \emph{interaction} \citep{bechinger2016active}. 
In fact, an advantage of the active matter concept is in its ``microscopic'' foundation, i.e. the ability to derive
equations for the macroscopic phase from the particle interaction \citep{bialke2013microscopic}.
This links the system dynamics and the dynamics of the system elements in a way that is both analytically tractable (on the macro level) and efficient to simulate (on the micro level).

Going from physical systems to animal societies or to online communities, a continuum model would not allow us to understand the system dynamics.
It simply does not address the rules by which the system elements have to interact, in order to collectively create the pattern we observe. 
With an increasing level of organization, also the internal degrees of freedom for the system elements increase.
To explain new \emph{experiments} in biological or social systems, e.g. about swarming behavior or online communication, requires us to take an \emph{agent-based approach} that matches the available \emph{data} about \emph{individual} interactions. 

Consequently, our research interest turns from continuous models to agent-based models.
The challenge then is to extent the dynamic framework applicable for non-animated systems to animated ones in a concise manner, which is the aim of this paper.

\paragraph{Active Brownian Particles.}

\emph{Active walkers} \citep{kayser-et-92,fs-lsg-94}, introduced in the early 1990's, were a first attempt to conceptualize this ``micro'' approach.
Similar to \emph{random walkers}, the dynamics of active walkers is influenced by random processes.
But additionally, they are capable of changing their environment, which is described by an environmental potential.
This constantly adapting potential then feeds back on the further motion of the walkers. 

\emph{Active Brownian Particles} were first introduced \citep{Schimansky-Geier1995}  as a generalization of active walkers, using modified Langevin equations.
The adaptive potential became a self-consistent adaptive field which, similar to a chemical substance, could decay and diffuse.
This extension allows to simulate structure formation in reaction-diffusion systems in a very efficient manner, using particle-based methods. 

The most important extension of this concept was to include the \emph{energetic conditions} that enable  Active Brownian Particles to become \emph{active}.
This was first done by considering \emph{negative friction} \citep{steuern-et-94}, a mechanism to pump energy into a system already discussed by Lord Rayleigh \citep{rayleigh-77}.
This energy was used to accelerate the Brownian particle, i.e. to allow for \emph{active motion}. 
A \emph{space dependent} negative friction coefficient models the take-up of energy only in certain areas, similar to the feeding places of animals. 
The drawback, however, was the instantaneous use of this energy.

This problem was cured by assuming that Active Brownian particles have an \emph{internal energy depot} \citep{fs-eb-tilch-98-let}.
Energy taken up from the environment is stored in this energy depot with a certain dissipation rate and can be transformed into kinetic energy, to allow for active motion.
This was a concise way to \emph{explain} the conditions for active biological motion, as opposed to \emph{postulating} a non-trivial velocity. 
Considering further the ability of Active Brownian Particles to generate a self-consistent adaptive field, this modeling approach has become the starting point to model a variety of biological phenomena, such as swarming and chemical communication.

Noteworthy, this concept has been already generalized 20 years ago as a framework for agent-based modeling \citep{fs-97-agent} with interdisciplinary applications \citep{schweitzer2007brownian}.
And the concept of Active Brownian Particles has also become a corner-stone in modeling \emph{active matter}, with \citep{bechinger2016active} or without \citep{ramaswamy2010mechanics} reference to these earlier investigations.

\paragraph{How to proceed.}

The aim of this paper is \emph{not} to present all details of the variety of models and applications discussed in the following. 
For these, we refer to the existing publications that also provide agent-based simulations and analytical investigations.
Our goal is rather to demonstrate that these different dynamics can be captured in a \emph{unifying} and \emph{overarching framework}.
The generalizing perspective taken here allows to \emph{highlight common principles} in the dynamics of active matter and to point out critical (energetic) conditions for the emergence of systemic properties.
With this, we provide an agent-based framework that, rarely enough, bridges between physical, biological and social phenomena.

\section{A dynamical framework for active matter}
\label{sec:matter}

\subsection{Non-equilibrium systems: Micro and macro perspective}
\label{sec:non-equil-syst}

Following the tradition of statistical physics, active matter can be described on two different levels.
On the macro level we focus on the system as a whole, to distinguish processes \emph{within} the system from exchange processes \emph{between} the system and its surrounding. 
From this perspective, active matter can be described as an \emph{open system} characterized by an influx of free energy (or matter, or information),
by internal dissipation and entropy production, and an outflux of energy with high entropy.
As such, active matter has the
properties of a \emph{non-equilibrium thermodynamical system} and the established methods can be applied. 
Hence, systemic properties, in particular thermodynamic functions and phase diagrams, have been investigated \citep{takatori2015towards}.

On the micro level we focus on the elements of the system and their interactions.  
From this perspective, active matter is composed of an ensemble of elementary units, denoted as \emph{agents}. 
These agents are \emph{active} in the sense that they can take up energy from the system, to use it for different activities.
Examples discussed in this paper include active motion, communication and other energy consuming types of interaction, or internal processes such as metabolism. 
The laws of thermodynamics then require an open system that can import and export energy/entropy, in accordance with the macrodynamics described above. 

In this paper, we mostly focus on the microlevel, i.e. on the agent-based approach. 
The dynamics of these agents are described by a set of stochastic equations which resemble the Langevin equation of Browian motion, therefore the notion of \emph{Brownian agents} has been established \citep{schweitzer2007brownian}.
The idea builds on Langevin's approach to explain the dynamics of a Brownian particle with velocity $v(t)$ by a superposition of two forces:
\begin{align}
  \label{eq:1}
  \frac{d v(t)}{d t} &= - \gamma_{v} v(t) + \sqrt{2S}\ {\xi(t)} 
\end{align}
The first term denotes the \emph{friction force} with $\gamma_{v}$ being the friction coefficient.
It essentially describes a \emph{dissipative process}, i.e. kinetic energy is decreased.
Hence, after some relaxation time expressed by $1/\gamma_{v}$ the Brownian particle would to come to rest, which is not observed under the microscope. 
To keep the particle moving, Langevin therefore assumed a \emph{stochastic force}, expressed in the second term. 
$\xi(t)$ is Gaussian white noise, i.e. it has the expectation value of zero and only delta-correlations in time:
\begin{align}
  \label{eq:2}
   \mean{\xi(t)}=0 \;; \quad \mean{\xi(t')\xi(t)}=\delta(t' -t) \;; \quad S=\gamma_{v} \frac{k_{B}T}{m}
\end{align}
$S$ denotes the strength of the stochastic force and is determined in physics by the  \emph{fluctuation-dissipation theorem}. 
$k_{B}$ is the Boltzmann constant, $T$ the temperature of the liquid the Brownian particle is immersed in and $m$ is the mass of the particle. 

For our further consideration, the \emph{structure} of Eq. \eqref{eq:1} is important.
The dynamics results from a \emph{superposition} of two different types of influences.
The first term denotes \emph{deterministic} forces which can be specified at the temporal and spatial scale of the agent.
This is the relaxation term, for the most simple case of a Brownian particle. 
The second term denotes \emph{stochastic} forces which summarize all  influences that are \emph{not} specified on these temporal and spatial scales. 
Of course, today it is known  that the Brownian particle keeps moving because of random collisions with molecules from the liquid too small to be observed in the microscope. 
So, in principle, on could \emph{derive} the second term from a more refined model.
But the ingenious idea here is to resist that temptation and instead proxy those unexplained influences by a random force as long as they do not exert a directed impact. 

To develop the dynamics of a Brownian particle into the dynamics of a Brownian \emph{agent}, this picture still misses \emph{interactions} between agents, \emph{internal degrees of freedom} to allow for different responses to the forces assumed, \emph{control parameters} to capture the influence of the environment. 
Noteworthy the picture also does not contain yet \emph{sources of energy} for activities that go beyond the level defined by the fluctuation-dissipation theorem. 
\emph{Active matter}, i.e. agents, if the micro perspective is considered, have the ability to perform certain \emph{activities}, ranging from directed motion to communication, from structure formation to collective excitations, which cannot be simply taken for granted. 
Hence, a microscopic model of active matter has to make the influx of energy explicit.

\subsection{Driving and driven variables}
\label{sec:driv-driv-vari}

Our micro level approach to active matter builds on an agent-based model that follows the concept of \emph{Brownian agents} discussed above.  
Precisely, in an ensemble of $N$ agents, each agent is described by two variables that follow a similar formal dynamics: 
\begin{align}
  \frac{d a(t)}{d t} &= - \gamma_{a}a(t) + \mathcal{G}_{a}(a,b,\mathbf{u}) + A_{a}{\xi_{a}(t)} 
  \label{eq:a} \\
  \frac{d b(t)}{d t} &= - \gamma_{b}b(t) + \mathcal{G}_{b}(a,b,\mathbf{u}) + A_{b}{\xi_{b}(t)} 
  \label{eq:b}
\end{align}
While the dynamics for the two variables is constructed in the same way, their meaning is very different. $a$ is the \emph{driving} variable, i.e. it describes the \emph{input} of energy and how this is related to different forms of \emph{activity}. $b$, on the other hand, is the \emph{driven variable} that describes the \emph{output} resulting from the use of energy. 

We will give a number of examples for the meaning of $a$, $b$ below. At this point, we mention that the damping factor $\gamma_{a}$, $\gamma_{b}$ ensures that, in the absense of any stimulus or external force, each of these variables will approach zero in the course of time. However, the additive stochastic term, $A{\xi}(t)$, that denotes the influence of random events, may prevent the relaxation toward zero. Different forms for this term are also discussed below.

The two functions $\mathcal{G}(a,b,\mathbf{u})$ eventually describe nonlinear couplings between the variables $a$ and $b$, where $\mathbf{u}$ represents a set of \emph{control parameters}. 
For their further specification, we use the following general ansatz:
\begin{align}
\label{eq:6}
\mathcal{G}_{a}(a,b,\mathbf{u}) & = \mathcal{F}_{a}(\cdot)\, \sum_{k=0}^n \alpha_k(b,\mathbf{u}) \ a^{k}(t) \\
  \label{eq:8}
      \mathcal{G}_{b}(a,b,\mathbf{u}) & = \mathcal{F}_{b}(\cdot)\, \sum_{k=0}^n \beta_k(a,\mathbf{u}) \ b^{k}(t) 
\end{align}
The power series should be seen as a general expression of \emph{hypotheses} about the nonlinear relation, examples of which are given later in the paper.  
Dependent on the application,  we will consider different orders of the power series where the coefficients $\alpha_{k}$, $\beta_{k}$ are determined by plausible arguments. 
The two free functions $\mathcal{F}_{a}(\cdot)$, $\mathcal{F}_{b}(\cdot)$ are left unspecified at the moment, we will use them later in introduce some non-local coupling between the variables $a$, $b$. 

To model the input of energy into the system, we are mostly interested in the dynamics of the driving variable $a$, for which we  discuss different representations. Importantly, the driving variable $a$ is usually assumed to relax fast compared to the driven variable $b$, i.e. following an adiabatic approximation we will in many cases describe $a$ by its  quasistationary equilibrium resulting from $\dot{a}\approx 0$: 
\begin{equation}
  \label{eq:a-adiab}
  a(t)=\frac{\mathcal{G}_{a}(a,b,\mathbf{u})+A_{a}{\xi_{a}(t)}}{\gamma_{a}}
\end{equation}

\section{Application:  Active motion}
\label{sec:appl-active-moti}

\subsection{Input of energy}
\label{sec:input-energy}

Let us start with the  most studied example in the context of this paper, namely active motion. 
It describes the ability of biological entities, from bacteria to fish and mammals, to move actively in a desired direction with a non-trivial velocity much larger than the thermal velocity, $v^{2} \sim k_{B}T/m$. 
Active motion plays a major role in models of swarming behavior \citep{Couzin2003,Romanczuk2009,chate2008collective}. 
However, many of these models just \emph{postulate} the non-trivial velocity, to focus on the interaction between agents.
They rarely discuss the energetic conditions for active motion. 

We have provided a model of active Brownian particles \citep{fs-eb-tilch-98-let} which takes these energetic conditions explicitely into account. 
Since in this model agents do not interact, we drop the agent index $i$ in the following. 
To cast this model in the framework of active matter, the driving variable $a(t)$ represents an \emph{internal energy depot} $e(t)$ of an agent, i.e. an \emph{internal degree of freedom}. 
Considering terms up to first order in $\mathcal{G}_{a}(a,b,\mathbf{u})$, the dynamics of the internal energy depot reads:
\begin{equation}
  \label{eq:depot2}
    \frac{d a(t)}{d t}\equiv \frac{d e(t)}{d t} = - \gamma_{e} e(t) + \Big\{\alpha_{0} + \alpha_{1}(v,\mathbf{u})\, e(t)\Big\} + A_{a}{\xi_{a}(t)} 
\end{equation}
The damping of the driving variable should model \emph{internal dissipation} of the energy depot at a rate $\gamma_{e}$.
The term $\alpha_{0}=q(r,t)$ considers the fact that the energy depot $e(t)$ can be filled up with a rate $q(r,t)$ that may explicitely vary with the location $r$ of the agent and with time $t$. 

The energy stored in the internal depot can be used for different activities. 
In the example at hand internal energy is converted into kinetic energy to propel the agent, with a rate proportional to the depot and a velocity dependent conversion function $d(v)$.
Hence, the \emph{driven} variable is the agent's velocity, $b\equiv v$, and $\alpha_{1}(v,\mathbf{u})=-d(v)$. 
For the conversion function, in the absence of empirically tested relations, we use again the ansatz of a power series:
\begin{equation}
  \label{eq:motion}
  d(v) = \sum_{k=0}^{n} d_{k}v^{k}(t)
\end{equation}
The constant term $d_{0}$, which yields $d_{0}e(t)$ in Eq. \eqref{eq:depot2} describes the dissipation of energy during the transformation of internal into kinetic energy. 
Hence, it makes sense to combine the two dissipative processes described by $\gamma_{e}$ and $d_{0}$ and to define the \emph{dissipation rate} $c=\gamma_{e}+d_{0}$.
$d_{1}=0$ because otherwise it would generate a bias toward positive/negative velocities \citep{Garcia2011d}. 
Hence, we are left with $d(v)=d_{2}v^{2}$.
So, Eq. (\ref{eq:a}) reads with $\alpha_{1}(v,d_{2})=-d_{2}v^{2}$, where the control parameter $d_{2}$ is the conversion factor:  
\begin{equation}
  \label{eq:depot}
\frac{d e(t)}{d t} = - \left[c+d_{2} v^{2}\right] e(t) + q(r,t) + A_{a}{\xi_{a}(t)} 
\end{equation}
We note that this way the dynamics of the internal energy depot has effectively become a \emph{balance equation}. 
The influx, or ``gain'', of energy is obviously given by the take-up of energy from the environment, $q(r,t)$, but also the stochastic force can result in a gain of energy if their mean value is \emph{not} zero, which is discussed further below. 
The outflux, or ``loss'', of energy results from dissipative processes during energy storage  and conversion ($c=\gamma_{e}+d_{0}$), but most importantly from the conversion of depot energy into kinetic energy for the movement of the agent. 
Before we further investigate the latter, we want to discuss different assumptions for  $q(r,t)$. 

\paragraph{Constant, fluctuating  or localized take-up of energy.}

The take-up function $q(r,t)$ can cover different cases for energy take up.
The most common one is to simply assume a \emph{constant} function $q(r,t)=q_{0}$ independent of location and time.
Such a constant take-up is reasonable for biological species that exist in a energy-rich environment which does not change fast.
For example, bacterial cells, \emph{Salmonella typhimurium}, have been used to test the take-up of energy in relation to the nutrition concentration of their environment \citep{Garcia2011d}.

Usually the take-up of energy may slightly fluctuate around a mean value $q_{0}$
\begin{equation}
q(r,t)=q_{0}\;; \quad \tilde{q}_{0}(t)=q_{0} + A_{a}{\xi_{a}(t)}
\label{eq:qtilde}
\end{equation}
where the stochastic process $\xi_{a}(t)$ is assumed to be Gaussian white noise with the properties given by Eq. \eqref{eq:2}. 
I.e., at any given time $\xi_{a}(t)$ is drawn from a normal distribution with mean zero and standard deviation $\sigma={A_{a}/\sqrt{2}}$, where $\sigma\ll q_{0}$.
Eq. \eqref{eq:depot} leads us in adiabatic approximation, Eq. \eqref{eq:a-adiab}, to: 
\begin{equation}
\label{eq:e-a}
  a(t)\equiv e(t)= \frac{\tilde{q}_{0}(t)}{c+d_{2}v^{2}(t)}\equiv e_{0}
\end{equation}
where $e_{0}$ is the \emph{quasistationary} value of the energy depot. 
It still depends on the actual velocity, but it is assumed that the energy depot relaxes very fast if $v(t)$ changes. 
A stationary value results only if also the velocity has reached a stationary value $v(t)\to v_{0}$  and white noise fluctuations of the energy take-up are neglected, $\tilde{q}_{0}(t) \to q_{0}$.

A more realistic assumption for the take-up of energy would depend on time and space,  to model for example localized food sources \citep{Ebeling1999}. Then, $q(r,t)=q_{0}$ holds only inside a spatial domain and the energy depot is only charged if the agent, during its motion, hits such a food source. 
This leads to an interesting \emph{intermittent} dynamics, i.e. a switch between periods of \emph{active} motion, as long as the depot is filled, and periods of \emph{passive} motion.
It is also related to \emph{bursty dynamics} \citep{karsai2018bursty}, dependent on the availability of energy.

\paragraph{Activity driven by shot noise.}

Instead of assuming a rather continous energy take-up of the agent, we can also model a stochastic process that increases the energy depot of the agent independent of the agent's active involvement \citep{Fiasconaro2009,Romanczuk2012a}. If the stochastic process is, for example, given by white shot noise, then we get  for the energy increase \citep{Strefler2009,czernik1997thermal}: 
\begin{equation}
  \label{eq:shot}
q(r,t)=0\;; \quad \hat{q}_{0}(t)=A_{a}{\xi_{a}(t)}=q\sum_{k=0}^{n(t)} \delta(t-t_{k}) 
\end{equation}
Different from $q_{0}$, which is the rate of energy take up, the parameter $q$ defines the amount of energy obtained in a pulse that occurs at each time step $t_{k}$. For white shot noise, the $t_{k}$ are the arrival times of a Poissonian counting process $n(t)$, i.e.  the probability that $n(t)=k$ such pulses occur in a time interval $(t-T,t)$ follows the  Poissonian distribution: 
\begin{equation}
  \label{eq:poisson}
\mathrm{Pr}\{n(t)=k\}=\frac{y^{k}}{k!}\exp{(-y)} \;; \quad y=\lambda T \end{equation}
Here $\lambda$ is the mean number of Dirac delta pulses per time unit, and $1/\lambda$ the average sojourn time between two delta pulses. Because the stochastic process $\xi_{a}(t)$, different from the above example, is the major source of energy, the average is not zero as in Eq. (\ref{eq:2}), but
\begin{equation}
  \label{eq:xiav}
  \mean{\xi_{a}(t)}=q\lambda \equiv q_{0}
\end{equation}
i.e. \emph{the averaged dynamics} of the energy depot results in the same quasistationary limit given by  Eq.  (\ref{eq:e-a}).

\subsection{Velocity}
\label{sec:velocity}

Assuming that the \emph{driving} variable describes the internal energy depot $e(t)$ of an agent, the conjugate \emph{driven} variable for active motion is given by its \emph{velocity}, $v(t)$, for which the dynamics reads: \begin{equation}
\label{Mdyn}
\frac{d b(t)}{d t} \equiv \frac{d v(t)}{d t} = - \gamma_{v} v(t)  + \bm{\Big\{}\beta_{0} + \beta_{1}(e,\mathbf{u})\, v(t) \Big\} + \sqrt{2S}\xi(t)
\end{equation}
This dynamics follows the already discussed Langevin Eq. \eqref{eq:1} for the velocity of a Brownian particle, with the addition of  the nonlinear function $\mathcal{G}_{b}(a,b,\mathbf{u})$ from Eq. \eqref{eq:b}, for which we consider terms up to first order.
One should note that the terms $\beta_{k}v^{k}$ all have the physical meaning of a force that is responsible for the acceleration/deceleration of the agent. 
Hence, we choose for $\beta_{0}= {F}(r)$, where the \emph{force} is assumed to result from an external potential, i.e. ${F}(r)=-\nabla U(r)$, or from interactions with other agents. 

The first-order term $\beta_{1}(e,\mathbf{u})v(t)$ describes the acceleration of the agent thanks to the energy taken from the internal depot $e(t)$. 
This term is of course is related to the term $\alpha_{1}(v,\mathbf{u}) e(t)$ describing the energy taken from the energy depot in the corresponding equation. 
To see how, let us assume that we consider kinetic energy $e=mv^{2}/2$ and set the mass $m=1$. 
Any change of the energy is then related to a change in velocity by $\dot{e}=v \dot{v}$. 
If only the first order terms are compared, this yields $\alpha_{1}e=v \beta_{1}v$ or: 
\begin{align}
  \label{eq:13}
  \beta_{1}(e,d_{2}) = \frac{\alpha_{1}(v,\mathbf{u}) e(t)}{v^{2}}= d_{2}e(t)
\end{align}
$\beta_{1}$ has the meaning of a \emph{negative friction}, i.e. it compensates, and can even exceed, the ``positive'' friction $\gamma_{v}$.
Negative friction was already discussed by Lord Rayleigh in his Theory of Sound \cite{rayleigh-77}. 
A violin bow transfers energy to the violin string by means of negative friction, which allows friction-pumped oscillations, often recognizable as a nice sound. 
In our case, the negative friction is responsible for the active motion of the agent with a non-trivial velocity, $v^{2}\gg k_{B}T/m$.
It results from the energy taken from the internal depot $e(t)$ and converted with a rate $d_{2}$ into kinetic energy, i.e. $\beta_{1}=d_{2}e(t)$. 
Hence, we find for the dynamics of the \emph{velocity} as the \emph{driven} variable: 
\begin{equation}
\label{Mdyn2}
\frac{d v(t)}{d t} = - \left[\gamma_{v} - d_{2} e\right] v(t)  -\nabla U(r)
+ \sqrt{2S}\xi(t)
\end{equation}
which is coupled to the driving variable $e$ via Eq. \eqref{eq:depot}.

\paragraph{Stationary states.}

The dynamics of the \emph{driven} variable, Eq. (\ref{Mdyn2}), and of the \emph{driving} variable, Eq. (\ref{eq:depot}), have to be solved together, to determine the stationary states. 
If we only consider a \emph{deterministic dynamics}, i.e. neglect the additive stochastic term for the moment, and further set ${F}(r)=0$, 
we have a set of coupled equations 
\begin{align}
\label{eq:3}
\dot{e}&=-\left[c+ d_{2}v^{2}\right]e +q_{0} \;;\quad \dot{v}=-\left[\gamma_{v} - d_{2}e\right] v    
\end{align}
with the stationary solution resulting from $\dot{e}=0$, $\dot{v}=0$ \citep{fs-eb-tilch-98-let,Ebeling1999,condat2002nem,Garcia2011d}:
\begin{equation}
\label{statvel}
v_0 = \pm \left({\frac{q_{0}}{\gamma_v} - \frac{c}{d_{2}}}\right)^{1/2} = \pm \left(\frac{c}{d_{2}}\right)^{1/2}\left(Q_{2} -1\right)^{1/2}
\;; \quad Q_{2}=\frac{q_{0}d_{2}}{\gamma_{v}c}
\end{equation}
This gives different insights with relevance for active matter: 

(i) We find a bifurcation dependent on the control parameter $Q_{2}$.
For $Q_{2}<1$, ${v}_{0}=0$ is the only (trivial) solution, for $Q_{2}>1$ non-trivial solutions ${v}_{0}\neq 0$ exist. 
The control parameter $Q_{2}$ precisely describes the relation between energy take-up $q_{0}$ and energy conversion $d_{2}$, on the one hand, and energy \emph{dissipation} from friction $\gamma_{v}$ and metabolism $c$, on the other hand.
I.e. we need a critical input of energy to observe active motion, or activity in general. 

(ii) Neglecting the decay of the driving variable, $- c e(t)$, which represents internal dissipation at the rate $c$, we do not find the bifurcation in the active behavior. 
Instead, we would observe continuous activity, albeit at different levels. 
Hence, the \emph{emergence}, i.e. the \emph{sudden} appearance of non-trivial phenomena, is inherently coupled to the existence of such dissipative processes. 

We add that 
the above discussion holds for the specific assumption $d(v)=d_{2}v^{2}$ used in Eq. \eqref{eq:motion}. 
Other possible assumptions for this conversion of internal into kinetic energy are discussed in \citep{Garcia2011d} and compared with experiments in bacteria.

\paragraph{External potential.}

The potential $U(r)$ in Eq. \eqref{Mdyn2} can be utilized to model different forces acting on the agent.  
A quadratic function, $U(r)=ar^{2}/2$, for example, could model a force toward the origin representing a ``home'' \citep{Ebeling1999,erdmann-et-00}.
Hence, the two location dependent functions $q(r,t)$ and $U(r)$ already allow to describe a rich environment for the agent with different ``food'' and ``nest'' locations. 

A linear function, $U(r)=ar$, on the other hand, results in a \emph{drift}, i.e.  the agent is forced to move into the negative direction, for example because of a current \citep{tilch-fs-eb-99}. 
We could identify the energetic conditions under which the agent can change its motion \emph{along} the gradient into a active motion \emph{against} the gradient. 
This further allowed us to obtain conditions for a steady current of agents actively moving in a \emph{ratchet potential} \citep{Fiasconaro2009,czernik1997thermal}. 
Interestingly, if stochastic forces are considered the net current of moving agents could be \emph{reversed} \citep{tilch-fs-eb-99}. 
Hence, including stochasticity into the description of active matter affects the dynamics not just quantitatively, but also \emph{qualitatively}.

\paragraph{Interaction potential.}

The choice of $\beta_{0}= {F}(r)$ is not restricted to forces resulting from an external potential.
We could as well consider forces resulting from interaction potentials, i.e. $F_{i}=\sum_{j}f_{ij}$ with $f_{ij}=\nabla U (r_{i},r_{j})$
where $r_{i}$ and are $r_{j}$ are the agents' positions. 
This extends the framework to include various models of \emph{collective motion} \citep{vicsek12:_collec,chate2008collective,Couzin2003} with relevance to biological systems.
For example, a long-range attraction potential can be used to control the spatial dispersion of a multi-agent system.
On the other hand, 
if the distance between two agents is below a certain threshold, a short-range repulsion potential ensures that they do not collide when moving, this way modeling avoidance behavior \citep{Mach2007}.
Combining various forces to model e.g. alignment or follow-the-leader behavior allows to model coherent swarming behavior of different species in a realistic manner.

\section{Application: Communication}
\label{sec:appl-comm}

\subsection{Communication field}
\label{sec:communication-field}

In the above examples, the energy take-up was always used for active motion, but not yet for interaction between agents which is another important ingredient of active matter. 
Interaction can be seen as a generalized form of communication \citep{schweitzer2007brownian}, even if that is not the conventional view in physics. 
Electrons generate an electrical field to ``communicate'', i.e. to provide information about their charge and location, to other electrons which can respond to this. 
Coming closer to the biological and social realm, agents ranging from bacteria, cells and insects to higher organisms, including humans, interact by means of communication. 
This implies to generate and to transmit a signal, but it also requires responses of other agents that receive and process this signal. 
Both, generating and responding to signals, come at a cost which is rarely explicitly considered in modeling approaches.
Our framework of active matter fills this gap, by considering that the energy take-up can be also used for communication. 

Let us first discuss the example of chemical communication. We assume that the agent produces a chemical marker at a rate $s(t)$ which requires energy, i.e. $s[e(t)]$.  
If the production of $s(t)$ is simply proportional to the internal energy depot, we find for the corresponding term in Eq. \eqref{eq:depot2}:
\begin{equation}
  \label{eq:chemical}
\alpha_{1}(v,\mathbf{u}) e(t) \equiv -\alpha e(t) = - s(t)
\end{equation}
Note that $\alpha$ is considered a constant here, i.e. there is no coupling to the velocity, but still to the driven variable via $s(t)$.
Combining the two energy consuming processes, internal dissipation at a rate $\gamma_{e}$ and production of a marker at a rate $\alpha$, we can now define $\hat{c}=\gamma_{e}+\alpha$.
Considering again  a continuous, slightly fluctuating take-up rate $\tilde{q}_{0}$, Eq. \eqref{eq:qtilde}, the quasistationary limit for the energy depot as the \emph{driving} variable reads: 
\begin{equation}
  \label{eq:e-a2}
  e(t) \to e_{0}= \frac{\tilde{q}_{0}}{\hat{c}}\;; \quad s_{0}=\alpha \mean{e_{0}}
\end{equation}
where the quasistationary production rate $s_{0}$ is proxied by a constant derived from the mean of $e_{0}$.

The \emph{driven} variable, in the communication scenario, is the \emph{communication field} $h(r,t)$ that is generated by the chemical markers produced by the agents. 
Assuming that these markers are continuously placed in the environment at the positions $r$ of the agents,  
the spatio-temporal communication field $h(r,t)$ aggregates these markers and defines their local concentration. 
In line with the general dynamics assumed for the driven variable, Eq. \eqref{eq:b}, the chemical concentration can decay over time at a rate $\gamma_{h}$. 
The additive stochastic term $A_{b}\xi_{b}$ is, on the aggregated level of the field, transformed into a diffusion term with $D_{h}=A_{b}^{2}/(2\gamma_{h}^{2})$ as the spatial diffusion coefficient of the chemical marker. 
This gives us the dynamics of the communication field as the \emph{driven} variable as follows \citep{fs-lsg-94}:
\begin{equation}
  \label{eq:hrt}
  \frac{\partial h(r,t)}{\partial t} = - \gamma_{h}h(r,t) + s_{0}\sum_{i=1}^{N}\delta\big[r-r_{i}(t)\big] + D_{h}\nabla h(r,t)
\end{equation}
The \emph{driving} variable, i.e. the energy depot, and the \emph{driven} variable, i.e. the communication field, are coupled  through $s_{0}$.  
The summation goes over all agents $i=1,...,N$ that produce the chemical marker with the quasistationary rate $s_{0}$ at their current position $r_{i}(t)$. 
I.e. from now on, we will use the agent index $i$ to refer to a larger number of agents \emph{interacting}.

\subsection{Biological aggregation}
\label{sec:biol-aggr}

We note that the dynamics of the communication field indeed captures essential processes involved in communication, such as \emph{writing}, i.e., the generation of information, \emph{dissemination}, i.e., the distribution of information as a diffusion process, but also a certain \emph{memory} effect. 
Generated information has a certain life time, and its value (novelty, importance) fades out over time. 

Only the \emph{impact} of the generated information is missing in this picture. 
Communication implies that there is also a \emph{reading} of these markers, and a certain type of \emph{response} to the signal which depends very much on the system under consideration. 
To illustrate this, we take first the example of biological aggregation. 
Different biological organisms from cells, to slime molds, amoebae and myxobacteria use a
chemical field to communicate. 
That means they generate chemical signals, but they also respond to these signals by changing their direction of motion dependent on gradients in the concentration.
This process is widely known as \emph{chemotaxis} \citep{keller-83}. 
We capture this by assuming the following equation of motion for the agents \citep{fs-lsg-94}: 
\begin{align}
  \label{eq:4}
  \frac{d {r}_{i}}{dt}={v}_{i} =
      \frac{\omega_{i}}{\gamma_{v}} \; \frac{\partial h({r},t)}{\partial{r}} +\sqrt{2 D}~
      \mathbf{\xi}_{i}(t)
\end{align}
This stochastic dynamics results from the already discussed Langevin equation in the overdamped limit, i.e. $\gamma_{v}$ is large and therefore the velocity becomes quasistationary. $D=S/\gamma_{v}^{2}$ is the spatial diffusion coefficient. 
The additional term is the deterministic force resulting from a gradient in the communication field. 
The agent \emph{reads} the information and \emph{responds} to it by preferably moving towards higher local concentrations of the chemical markers, with $\omega_{i}$ as the (individual) sensitivity. 

Our model of active matter now includes a \emph{feedback loop}: agents produce chemical markers at their current position, this way establishing the communication field, but the communication field feeds back to the movement of the agents. 
This leads to a local amplification: agents reinforce higher concentrations of markers because they preferably go there. 
The local reinforcement is counterbalanced by two processes, (i) the decay and (ii) the diffusion of information. 
The latter can be seen as a long-range coupling of the distributed activities of agents. 
The decay, on the other hand, ensures that all information that is not reinforced will disappear over time. 
This induces a \emph{competition process}, where local maxima of chemical markers compete for the agents to be maintained.

This model has direct applications in biological aggregation.
For example, larvae of the bark beetle \emph{Dendroctonus micans}, after hatching, need to gather and to form clusters, to defeat some poison emitted by their host plant \citep{deneubourg-et-90}.
This is done by means of chemotactic communication. 
In a first stage, larvae form small local clusters by following the chemical gradient.
In a second stage, these local clusters start to compete, i.e. their further growth is at the expense of other clusters disappearing, a process known as Ostwald ripening in physics \citep{12-schmelzer-fs-87-b}. 
This way, eventually all larvae meet in one large cluster.

The dynamics of this aggregation process can be described on two levels, the micro level of communicating agents which is investigated by means of agent-based stochastic computer simulations, and the macro level of \emph{distribution functions} analysed mathematically by means of two coupled differential equations. 
In the current case, these distributions represent (i) the chemical concentration and (ii) the spatial density of agents. 
Different adiabatic approximations then allow to formally derive selection equations for the competing clusters or effective diffusion equations for the agent density \citep{fs-lsg-94}. 
We note again that the dissipation, i.e. the decay of information, plays a crucial role in the emergence of structures in active matter, which is the formation of clusters in our case.

\subsection{Self-assembling networks}
\label{sec:self-assembl-netw}

The basic communication model described above can be extended to describe the self\-assembing/self\-wiring of networks, as it was found e.g. in neuronal networks \citep{segev-acs-98}. 
Networks consists of nodes and links to connect them. 
The terms \emph{self\-assembing/self\-wiring} refer to the fact that links between nodes cannot simply be drawn as lines, but have to be \emph{physically} created in active matter. 

To model this by means of our agent-based framework \citep{fs-tilch-02-a}, we consider two different kinds of nodes distinguished by the index $-1,+1$. 
These nodes are spatially distributed, their positions denoted as $r_{j}^{z}$. 
Agents, while moving, first have to discover these nodes and then have to connect nodes of opposite type, i.e. ``$-$'' nodes are connected to ``$+$'' nodes, and vice versa.
Each agent is characterized by a discrete internal degree of freedom $\theta_{i}\in\{-1,+1\}$, which is changed only if the agent hits a node and then takes the value of the node index, i.e. $-1$ or $+1$.
Once an agent hits a node, its energy depot is charged to a maximum value $e_{\mathrm{max}}$. 
Hence, the take-up of energy  is \emph{not} assumed to be constant in space and time. Instead:
\begin{equation}
  \label{eq:qrt}
  q_{i}(r_{i},t)=\big[e_{\mathrm{max}}-e_{i}(t)\big]\;\delta\big[r_{j}^{z}-r_{i}(t)\big]\;\delta\big(t-t^{z}_{i}\big)
\end{equation}
where $t^{z}_{i}$ is the time when the agent $i$ hits one of the nodes $z$.

With Eq. \eqref{eq:chemical} the dynamics of the energy depot reads in the deterministic limit:
\begin{align}
  \label{eq:7}
\frac{d e_{i}(t)}{dt}= - \hat{c} e_{i} +   q_{i}(r_{i},t) 
\end{align}
As before, the energy from the depot is used to generate information, e.g. chemical markers.
In this application, instead of one diffusing chemical \emph{two} different \emph{non-diffusive} markers are produced. 
The new element is a state dependent production rate $s_i(\theta_i,t)$, 
\begin{equation}
s_i(\theta_i,t)=s_{\mathrm{max}}\,\frac{\theta_i}{2}\Big[(1+\theta_i)\,
\exp\{-\hat{c}\,(t-t_{i}^{n+})\}\\  
-\,(1-\theta_i)\,\exp\{-\hat{c}\,(t-t_{i}^{n-})\}\Big]
\label{prod}
\end{equation}
with $s_{\mathrm{max}}={\alpha} e_{\mathrm{max}}$. 
$t_{n+}^{i}$, $t_{n-}^{i}$ are the times when agent $i$ hits any of the nodes $+1$ or $-1$.
The \emph{time dependent} production rate results from the fact that the internal energy depot of the agent is \emph{not} in a quasi-stationary equilibrium, i.e. $s(t)\propto e(t)$ as given by Eqn. \eqref{eq:chemical}. 

The \emph{driven} variable is again the communication field, which now has two components to reflect the two different information, $\{-1,+1\}$.
Because information does not diffuse here, the dynamics of the communication field reads:
 \begin{equation}
\frac{\partial h_{\theta}({r},t)}{\partial t}=-\gamma_{h}\,
h_{\theta}({r},t) + \sum_{i=1}^{N} 
s_i(\theta_i,t)\;\delta_{\theta;\theta_{i}}\;
\delta\big[{r}-{r}_i(t)\big] \label{h-net-nd}
\end{equation}
In order to close the feedback loop of communication, we have to specify how agents respond to the information from the communication field. 
Here, we use Eq. \eqref{eq:4} again, i.e. agents respond to the gradient, but because there are two different fields $-1,+1$, our main assumption is that agents only pay attention to the component of the field they currently \emph{not} produce. 
That means agents departing from a ``$+$'' node, generate a marker $+1$ at an exponentially decaying rate, but in their movement they are guided by the gradient resulting from the field component $-1$.
This gradient, by construction, guides them to one of the ``$-$'' nodes.
Arriving there, agents switch their internal state to $-1$, start creating the marker $-1$, but follow the gradient from component $+1$, and vice versa. 

This rather simple feedback mechanism determines agents to ``weave'' connetions between nodes of opposite sign \citep{fs-tilch-02-a}. 
Of course, established links provide a screening effect in the neighborhood, therefore the link density becomes saturated over time. 
Dependent on control parameters such as the spatial density of agents, $N/A$, the relative production $s_{\mathrm{max}}/\gamma_{h}$ and the diffusion constant $D$, the agents are able to discover and to link all nodes in the system that are within a critical distance.
I.e. in the optimal range of parameters, the connectivity of the system is high and the spatial distribution of ``$+$'' and ``$-$'' nodes can be used to control the topology of the resulting networks, to obtain e.g. lattices or hub-spoke structures \citep{fs-tilch-02-a}. 
  
This model of structure formation in active matter has its application in the self-wiring of neural structures. 
A neuron grows, for example, from the retina of the eye towards the optic tectum (or
superior colliculus) of the brain, without ``knowing'' from the outset
about its destination node in the brain. 
It is known that gradients of different chemical cues play a
considerable role in this navigation process. They provide a kind of
\emph{positional information} for the navigation of the growth cones
\citep{tessier-goodman-96}. But in the very beginning, this positional information
has to be generated interactively, and only in later stages may lead to
established pathways.   

The model can be also applied to the formation of trails in ants \citep{Schweitzer.Lao.ea1997Activerandomwalkers}. 
The two different kinds of nodes are then the nest ($-1$) and the food sources ($+1$).
Starting from the nest, ants have to discover the food sources and then link them back to the nest, for exploitation. 
The success information is only produced after ants have discovered the food, and will lead other ants to that area. 
The information generated by the ants leaving the nest, on the other hand, is utilized by the successful ants to return to the nest. 
This model becomes more realistic by using refined assumptions, such as a success dependent sensitivity expressed by the strength of the stochastic force, or different rules for scouts and recruits \citep{Schweitzer.Lao.ea1997Activerandomwalkers}
We note that this model was also applied to the  formation of trail system by pedestrians \citep{helbing-fs-et-97}. 
Hence, the notion of \emph{active matter} can be expanded also to biological and social systems. 

\subsection{Human online chats}
\label{sec:human-activ-patt}

The framework of active matter can be also applied to human communication, for example in online chatrooms \cite{Garas2012}. 
Again, the \emph{driven} variable is the communication field which now consists of only one component and  updates instantaneously, without diffusion.  
It could be imagined as a computer screen that displays all messages from a chat within a certain time interval. 
New messages arrive at the top of the screen and make the highest impact, whereas older messages move down to the bottom as new messages come in, this way becoming less influential.
So the screen is always updated and information can fade out. 

The difference to the above communication examples is in the specification of the \emph{driving} variable, which is the energy depot.
Before, it was assumed that energy from the depot is available either constantly, as in the case of biological aggregation, or continuously but with a decaying amount, as in the case of self-assembling networks. 
This availability of energy then directly determines the amount of information produced by the agent, Eq. \eqref{eq:chemical}.

For human communication, both offline \citep{oliveira2005human} and online \citep{Garas2012}, we know that the inter-activity time $\tau$ between two communication acts of a given person follows a power-law distribution: 
\begin{align}
  \label{eq:5}
  P(\tau)\propto \tau^{-\kappa}\;;\quad \kappa=3/2
\end{align}
That means in an agent-based model the agent produces a constant amount of information $s_{0}(t)$ only at times that are defined by the sequence of $\tau$ values. 
This can be ensured by an energy depot that has a nonzero, but constant, value $e_{0}(t)$ only at these respective times, and is zero otherwise to prevent writing. 
To cope with this, the driving variable can in adiabatic approximation be written as: 
\begin{equation}
  \label{eq:e-comm}
  a(t)\equiv e_{0}(t)= \frac{{q}}{\hat{c}} \sum_{k=0}^{n(t)} \delta(t-\tau_{k})  
\end{equation}
where $\tau_{k}$ are realizations drawn from the interevent time distribution $P(\tau)$, Eq. \eqref{eq:5}, which is different the from the Poissonian distribution for shot noise, Eq. \eqref{eq:poisson}.
But still, at each of these arrival times the depot is filled with a rate $q$, which then allows the immediate generation of information. 

Noteworthy, a mean inter-activity time $\mean{\tau}$ is in general only defined if we assume a minimum and maximum inter-activity time $\tau_{\mathrm{min}}$, $\tau_{\mathrm{max}}$. 
Then
\begin{equation}
  \label{eq:taumin}
  \mean{\tau}=\frac{\tau_{\mathrm{max}}^{2-\kappa}- \tau_{\mathrm{min}}^{2-\kappa}}{\tau_{\mathrm{max}}^{1-\kappa}- \tau_{\mathrm{min}}^{1-\kappa}}
\end{equation}
and we can again calculate an average take-up rate $q_{0}=q/\mean{\tau}$ for the energy depot which brings us back to Eq. \eqref{eq:e-a2}.

\section{Application: Emotional Influence}
\label{sec:appl-emot-infl}

\subsection{Valence and arousal}
\label{sec:valence-arousal}

So far, we have always considered the internal energy depot as the \emph{driving} variable, whereas the \emph{driven} variable was either the velocity, as in the case of active motion, or the communication field, as in the case of interacting agents. 
Now, we keep the communication field as the \emph{driven} variable, but use different assumptions for the \emph{driving} variable, to model emotional influence. 

For this, we consider that the information generated by the agent is no longer just a function of the internal energy depot, $s_{i}[e_{i}(t)]$, Eq. \eqref{eq:chemical}, but a function of two different driving variables, valence, $x_{i}(t)$, and arousal, $y_{i}(t)$, i.e. $s_{i}[x_{i}(t),y_{i}(t)]$. 
According to the so-called circumplex model \citep{Russell1980}, 
the two variables are used to quantify \emph{emotions},  $\mathcal{E}(x,y)$, 
by a position in the two-dimensional $(x,y)$ space. 

\emph{Valence} $x$ refers to the pleasure associated with an emotion, while \emph{arousal} $y$ refers to the degree of activity induced by the emotion. 
Both are normalized to a range $(-1,+1)$ and can be measured in different ways.
Of interest in our context is the \emph{sentiment analysis} which allows to extract emotions from written text. 
I.e. the underlying assumption, as before, is a communication between agents, e.g. in an online setting (chats, fora) via the exchange of text messages. 
These text pieces, in addition to some factual information, also contain emotions that affect other agents when reading the text. 
They may then respond by writing a reply, this way expressing their emotions. 
If we abstract from the text, this interaction describes an emotional influence of agents. 

For an individual, changing emotions generate a trajectory in the $(x,y)$ plane characterized by a large amount of noise \citep{Kuppens2010}.
Therefore, it is reasonable to assume for \emph{both} driving variables the stochastic dynamics proposed in Eq. \eqref{eq:a}.
It remains to specify the nonlinear function, $\mathcal{G}_{a}$, for which we start with the  general ansatz of Eq. \eqref{eq:6}. 
This time, however, we make use of the free functions $\mathcal{F}_{a}(\cdot)$ by choosing 
 \begin{align}
 \label{eq:66}
\mathcal{F}_{x}(\cdot) = h_{\pm}(t) \;;\quad \mathcal{F}_{y}(\cdot)= \hat{h}(t)= h_{+}(t)+h_{-}(t)     
 \end{align}
That means, the dynamics of both valence $x$ and arousal $y$ are expressed by the power series further discussed below, but there are noticeable differences in the assumed dependence on the existing emotional information, expressed by $\mathcal{F}_{x}(\cdot)$ and $\mathcal{F}_{y}(\cdot)$. 
Both free functions depend on the respective communication field $h$, which is the \emph{driven} variable.
Because of positive and negative emotions, we assume that their information is aggregated in two different components of the communication field, $h_{+}$ and $h_{-}$. 
Both components follow the same dynamics given by Eq. \eqref{h-net-nd}, i.e. their value decays with a rate $\gamma_{h}$ but is increased by the emotional information produced by the agents, $s_{i}[x_{i}(t),y_{i}(t)]$.
Additionally, we can also consider input from external events, $I_{\pm}(t)$, that increase the value of the positive or negative  emotional information.
To further specify the production rate of emotional information, we choose:
\begin{align}
   s_i[x_{i}(t),y_{i}(t)] = f[x_i(t)]\, \Theta[y_{i}(t) - \mathfrak{T}_i]
\label{eq:9}
  \end{align}
$\Theta[z]$ is the Heavyside function, which is 1 if $z\geq 0$ and 0 otherwise. 
It means that agent $i$ produces emotional information only if its arousal $y_{i}$ reaches the level of the individual threshold $\mathfrak{T}_{i}$.
Whether this information contains positive or negative emotions is decided by the agent's valence $x_{i}(t)$. 
The function $f[x_{i}(t)]$ can for simplicity just distinguish between positive and negative emotions, but other dependencies are also possible. 

Once the communication field is established by the emotional expressions, it feeds back on the driving variables as specified in Eqs. \eqref{eq:6}, \eqref{eq:8}.
Here we assume that valence is only affected by the emotional information that matches the agent's emotion, i.e. by $h_{-}$ for agents with negative valence \emph{or} by $h_{+}$ for agents with positive valence.
For arousal, on the other hand, it is considered that both positive \emph{and} negative emotional information increases the arousal of the agent, hence the information is additive. 
Importantly, if the arousal reaches the individual threshold $\mathfrak{T}_{i}$,  $y_{i}$ is set back to zero. 

\subsection{Nonlinear feedback}
\label{sec:nonlinear-feedback}

To complete the feedback cycle, we have to further specify the coefficients of the power series in $\mathcal{G}_{a}(a,b,\mathbf{u})$,  Eq. \eqref{eq:6}.
For the dynamics of \emph{valence} we find \citep{Schweitzer2010a}:
\begin{align}
  \label{eq:11}
\frac{d x_{i}(t)}{dt}  
 & = - \gamma_x x_{i}(t) +
   \Big\{ h_{\pm}(t)\left[\alpha_{0x}+\alpha_{1x}x_{i}(t)+\alpha_{2x}x_{i}^{2}(t)+\alpha_{3x}x_{i}^{3}(t) \right] \Big\} + A_{x}\xi_{x}(t) \nonumber \\ 
& = - \gamma_x x_{i}(t) + x_{i}(t){h}_{\pm}(t) \Big\{\alpha_{1x} + \alpha_{3x}x_{i}^{2}(t)\Big\} + A_{x}\xi_{x}(t)
\end{align}
Here, we have considered contributions up to 3rd order in $x$. 
To  allow for a ``silent'' mode $x(t)\to 0$,  $\alpha_{0x}=0$ has to be chosen.
Further, in order to treat  positive and negative valences as ``equal'' and to not introduce a bias, we have to set $\alpha_{2x}=0$. 
The analysis of the remaining equation \citep{Schweitzer2010a} then tells that non-trivial solutions $x\neq 0$ require $\alpha_{1x} \cdot h_{\pm} > \gamma_{x}$. 
In this case, \emph{collective emotions} can emerge which involve all agents. 
We note again that emergence, i.e. the \emph{sudden appearence} of collective states, occurs only if dissipation ($\gamma_{x}$) is involved. 

To obtain \emph{activity}, a regime with high (positive or negative) valence and high emotional information $h_{\pm}$ is required. 
But this regime can only occur if arousal is high enough to generate some emotional expression $s_{i}$ in the first place.  
We obtain from Eqs.  \eqref{eq:a}, \eqref{eq:8}:
\begin{align}
  \label{eq:12}
\frac{d y_{i}(t)}{dt} & = - \gamma_y y_{i}(t) +
    \hat{h}(t)\Big\{\alpha_{0y}+\alpha_{1y}y_{i}(t)+\alpha_{2y}y_{i}^{2}(t) \Big\}  + A_{y}\xi_{y}(t)
\end{align}
Here, $\mathcal{G}_{a}(a,b,\mathbf{u})$ considers contributions up to 2nd order in $a$.
A positive arousal requires the coefficient $\alpha_{0y}>0$, which can be seen as analogy to the constant take up of energy $q_{0}$ in the case of the internal energy depot. 
The two other coefficients $\alpha_{1y}$, $\alpha_{2y}$ are responsible for the nonlinear self-reinforcement of arousal, thus $\alpha_{1y}\neq 0$ should be chosen  \citep{Schweitzer2010a}. 
The crucial coefficient is $\alpha_{2y}$ because it decides about the long-term dynamics possible.
If $\alpha_{2y}<0$, the arousal dynamics becomes saturated. 
If this saturation level is above the individual threshold $\mathcal{T}_{i}$, the agent will generate an emotional expression. 
After that $y_{i}$ is set back to zero. 
If fluctuations are included these may then push the agent's arousal to negative values, from which it will not return to positive arousal. 
Hence, we obtain a scenario where agents express their emotions most likely only once. 
This may lead to collective emotions, but not repeatedly.

The situation changes if we consider $\alpha_{2y}>0$. 
Then, instead of a saturated dynamics, we may obtain two different stationary solutions with negative arousal. 
At low levels of emotional information $\hat{h}(t)$, e.g. no generation of emotional information, fluctuations are able to push the agent's arousal to positive values, from where a new communication cycle starts.
Hence, we obtain a scenario with \emph{waves} of collective emotions over time.

To summarize the dynamics of emotional influence, stochastic fluctuations are in fact crucial to reach an active regime. 
They first push agents to a positive arousal which is then amplified by the positive feedback, until it reaches the threshold. 
This then generates emotional expressions that establish a communication field which in turn feeds back on the agent's valence and arousal. 
While valence, in this dynamics, is responsible for the ``content'', i.e. the sign of the emotional information generated, arousal decides about the activity pattern. 
But it is worth noticing that arousal does not drive valence, consequently both are the \emph{driving} variables, whereas the emotional expressions and the components of the resulting communication field are the \emph{driven} variables. 

Despite the rather abstract description, this model has performed remarkably well in reproducing significant features of emotional online communication, such as the emotional persistence of users, activity patterns of users dependent on emotional stimuli, or
the emergence of collective emotions \citep{Garas2012,Schweitzer2016,Schweitzer2015}.

\subsection{Emotions driving opinions}
\label{sec:appl-opin-dynam}

As the last application, we consider the case that the driving variables, valence and arousal, drive \emph{opinions} through  the production of information, $s_{i}[y_{i}(t),v_{i}(t)]$.
This information results again in a communication field that influences opinions, i.e. opinions are driven by underlying emotions. 
Taking for example political systems \citep{Abisheva2015}, the frequently observed strong polarization in opinions can hardly be explained just by rational or utility-based arguments. 
Instead, the emergence of polarization in a heated political climate is rather due to ``irrational'' processes fast enough to impact opinions beyond control. 

Emotions are defined as \emph{short-lived psychological states}, hence their dynamics relaxes fast compared to the dynamics of opinions and we can indeed separate the time scales of these two dynamics. 
For the dynamics of emotions, we utilize Eqs. \eqref{eq:11}, \eqref{eq:12}. 
For the dynamics of the \emph{opinion} $\theta_{i}(t)$, which is the \emph{driven variable} here, we assume in accordance with Eqs. \eqref{eq:b}, \eqref{eq:8} \citep{schweitzer18}: 
\begin{align}
 \label{eq:va-basic}
        \frac{d\theta_i(t)}{dt} &= - \gamma_{\theta }\, \theta_i(t) 
+\Big\{\beta_{0}+\beta_{1}\theta_{i}(t)
+\beta_{2}\theta_{i}^{2}(t)+\beta_{3}\theta^{3}_{i}(t)\Big\}
+ A_{\theta i}\xi_{\theta}(t)
    \end{align}
Here $\theta_{i}\in (-1,+1)$ denote \emph{continuous} opinions where negative values may indicate left-wing and positive values right-wing positions. 
Their values are not necessarily bound to the range given, even more extreme opinions may be possible but not frequent. 
Again, for the dynamics we consider contributions of the power series in $\mathcal{G}_{b}(a,b,\mathbf{u})$ up to 3rd order. 
The higher-order terms are useful to encode subtleties in the opinion formation, e.g. 
 $\beta_{2} \neq 0$ would account for a global bias toward left/right opinions.
$\beta_{3}<0$, on the other hand, indicates a common preference for consensus. 
If $\abs{\beta_{3}}$ is large, this favors consensus, whereas small $\abs{\beta_{3}}$ favor polarization \citep{schweitzer18}.

The coupling between the \emph{driving} variables $x_{i}(t)$, $y_{i}(t)$, and the \emph{driven} variable $\theta_{i}(t)$ is effectively provided via the communication field $h$ with its two components $h_{+}$ and $h_{-}$, on which the coefficients $\beta_{k}$ depend.
The important coefficients are $\beta_{0}$ and $\beta_{1}$. 
Obviously, if $\beta_{0}=0$ and no bias $\beta_{2}$ is considered, $\beta_{1}$ already decides whether there is only one (trivial) opinion in the long run ($\beta_{1}<\gamma_{\theta}$) or whether there could be (a coexistence of) two different opinions $\theta\neq 0$ ($\beta_{1}>\gamma_{\theta}$).

Hence, what matters for the term $(\beta_{1}-\gamma_{\theta})\theta_{i}$ is the \emph{difference} between $\beta_{1}$ and $\gamma_{\theta}$.
A coexistence of two different opinions, i.e. \emph{polarization}, should occur in a regime with high emotional information, $\hat{h}(t)=h_{+}(t)+h_{-}(t)$.
We recall that the communication field $\hat{h}(t)$ measures the activity resulting from the emotions.
So it makes sense to set $(\beta_{1}-\gamma_{\theta})$ equal to $(\hat{h}(t)-\hat{h}_{\mathrm{base}})$.
I.e., $\hat{h}(t)$ has to overcome some threshold value $\hat{h}_{\mathrm{base}}$ in order obtain nontrivial opinions and polarization.

Eventually, we have to consider that both left- and right-wing opinions, even if they coexist, may be present in the system with  different frequency. 
This is decided by the coefficient $\beta_{0}$ for which we choose $\beta_{0}(x,h)= -w\,\bar{x}(t)\, \hat{h}(t)$ \citep{schweitzer18}. 
$w$ is a dimensional constant. 
Note that $\bar{x}(t)=[h_{+}(t)-h_{-}(t)]$ is a measure of the average valence in the system which,
in the simplest case, can be expressed by the difference in the available positive and negative emotional information.

If $\beta_{0}>0$ and $\abs{\beta_{0}}$ is large, i.e. if we have a situation with high emotional information which is mostly negative, then the corresponding opinions are also mostly ``negative''. 
On the other hand, if  $\beta_{0}<0$ and $\abs{\beta_{0}}$ is large, the corresponding opinions are mostly ``positive''. 
The interesting case is for intermediate values of $\abs{\beta_{0}}$ because they allow for the nonstationary coexistence of left/right opinions dependent on the emotional response of agents. 
I.e., this is the parameter range where polarization of opinions emerges based on emotional interactions.

\section{Discussion}
\label{conclusions}

Active matter is a recent concept to describe the active motion and the formation of coherent structures in systems 
with the ability to take up, and to convert, energy.
As such, active matter shares features of \emph{self-organization}, which is found not only in physical systems, but also in biological and in social systems.
This raises the question about an overarching framework that allows to model active matter on these very different levels of organizations by using the same principles.
Macroscopic approaches have proven useful to model, for instance, the hydrodynamics of active matter \citep{julicher2018hydrodynamic,marchetti2013hydrodynamics,Toner2005}, but cannot be easily extended to model swarming behavior of animals or social communication \citep{Couzin2003}. 
Therefore, ``microscopic'', i.e. \emph{agent-based}, approaches are more promising to capture the complex interactions in such systems.

But can we employ the \emph{same} agent-based approach for these very \emph{different} systems, or do we need specific agent-based models for each of them?
This problem is addressed in this paper.
We demonstrate that there are indeed unifying dynamic principles that can be utilized to model physical, biological and social systems as active matter.

Our agent-based framework is characterized by the following features:

\emph{(i) Distinction between driving and driven variables: \ }
Driving variables allow  to model the \emph{take up}, storage and conversion of \emph{energy}  which is crucial to model subsequent \emph{activites}, whereas driven variables describe what the energy is used for. 
In this paper, we have discussed as application scenarios three different activities: (a) active motion (Section 3), (b) communication (Section 4) and (c) emotional influence (Section 5). 
(a) and (b) can be widely found on different levels of biological organization. 
Both cases have the same \emph{driving} variable, the energy depot, whereas the \emph{driven} variable is in (a) the agent's velocity and in (b) the communication field to model indirect interaction.
(b) and (c) can be observed on different levels of biological and social organizations.
Both cases share the same \emph{driven} variable, i.e. a two-component communication field, but the \emph{driving} variable is in (b) the agent's energy depot and in (c) the agent's emotion characterized by valence and arousal.

\emph{(ii) Stochastic dynamics and nonlinear feedback: \ }
For both driving and driven variables we have assumed the same kind of stochastic dynamics that resembles the Langevin equation of Brownian motion, Eqs. \eqref{eq:a}, \eqref{eq:b}. 
This contains a damping term for relaxation and an additive  stochastic force. 
The new element is the additional term in each equation to describe the nonlinear feedback between \emph{driving} and \emph{driven} variable. 
These non-linear functions $\mathcal{G}_{a}(a,b,\mathbf{u})$, $\mathcal{G}_{b}(a,b,\mathbf{u})$ are introduced by means of a power series, Eqs. \eqref{eq:6}, \eqref{eq:8}, where the coefficients $\alpha_{k}$, $\beta_{k}$ are in fact functions that depend on the driving and driven variables and on a number of control parameters $\mathbf{u}$. 

\emph{(iii) Analytical and empirical assessment: \ }
The non-linear functions allow us to encode \emph{testable hypotheses} about the relation between driving and driven variables.
Such tests can be performed mathematically, in particular applying bifurcation and stability analyses.  These reveal the conditions under which a certain dynamic behavior of the system can be expected \citep{Schweitzer.Lao.ea1997Activerandomwalkers,fs-tilch-02-a,Schweitzer2010a,ebeling2015unlikely,tilch-fs-eb-99,Ebeling1999}.
Moreover, the energetic conditions for active motion, self-assembling of networks, swarming behavior or collective emotions have been determined.
The role of dissipation in obtaining \emph{different} dynamical regimes could be highlighted. 
But these hypotheses can be also tested in experiments \citep{Birbaumer2011a,Garcia2011d,Schweitzer2016,Garas2012} which allow to calibrate the respective parameters.

\emph{(iv) Formal relation between micro and macro description: \ }
Using methods of statistical physics, the framework of Brownian agents \citep{schweitzer2007brownian} was developed such that it allows to formally derive the \emph{systemic} dynamics from the dynamics of the system elements, at least in some approximation (mean-field assumptions, separation of time scales).
This is a considerable advantage in comparison to other agent-based approaches that only rely on freely defined rules and extensive computer simulations.
This advantage takes effect also for modeling active matter in the broader sense used in this paper. 
That means, in addition to efficient agent-based simulations that already include stochastic influences, we are able to project the systemic properties, to estimate conditions under which collective behavior and structure formation emerge.

While comparable methods for physical systems already exist for long, the major advantage of our framework, as pointed out in this paper, is its applicability to a variety of non-physical systems, in particular biological and social systems.
To bridge a modeling approach between non-animated and animated systems is quite rare and could be highly criticized for fundamental reasons.
But such arguments can already be refuted because it was demonstrated that seemingly ``social'' phenomena, such as collective emotions in online social media, can be remarkably well described using this framework -- not in an abstract manner, but based on data-driven model calibration.

\end{document}